\documentclass[12pt]{article}
\newcommand{\mfrac}[2]
{\mbox{\footnotesize$\displaystyle\frac{\raisebox{-0.1em}{\mbox{$\,#1\,$}}}
{\,#2\,}$}}

\begin{document}

\centerline{\Large On $\Psi$-function for finite-gap potentials}
\vskip 0.5cm
\centerline{\large {Ustinov N.V.,\ \  Brezhnev Yu.V.}}
\vskip 0.5cm
\centerline{\it {\small Kaliningrad State University}}

\centerline{\it \small e-mail: \ {\tt n\_ustinov@mail.ru, \ brezhnev@mail.ru}}
\vskip 0.5cm

\noindent
Being nonconstructive in general case $\Theta$-functional formulae for
solutions of the spectral problems and for finite-gap potentials leave open a
question on explicit representation of \mbox{$\Psi$-function} in terms of the
potentials.
A connection between spectral problem
\begin{equation}\label{Sch}
\Psi''-u(x)\,\Psi=\lambda\,\Psi
\end{equation}
and stationary solutions of higher equations of KdV hierarchy in a case, when
a spectrum of operator (\ref{Sch}) has finite number of prohibited zones, was
established in work \cite{novikov}.
Next, a notion of finite-gap potentials was significantly generalized.
They are considered now as potentials, whose $\Psi$-function is Baker-Akhiezer
function on the algebraic curve of finite genus.
In this report we show that initial stationary interpretation leads naturally
to the expressions for $\Psi$-function in terms of arbitrary finite-gap
potential and for the algebraic curve in terms of first coefficients of its
polar expansion.

An operator is finite-gap if there exists the operator bundle commuting with
them \cite{krichever}.
To construct it, one can invoke evolution equation and impose stationary
reduction.
Like that for Eq.(\ref{Sch}), solving well-known $A$-equation of Lax pair by
separation of variables, we obtain a desired bundle:
\begin{equation}\label{tmp}
F(\lambda;u,u',\ldots)\,\Psi'-\mfrac12\,F'(\lambda;u,u',\ldots)\,\Psi=
\mu\,\Psi,
\end{equation}
where $\mu$ is a parameter of the variables separation.
Integrating, one results in expression
\begin{equation}\label{psi}
\Psi(x;\lambda)=\sqrt{F\,}\,\exp\!\int\!\mfrac{\mu}{F}\,
\mbox{\footnotesize $dx$},
\end{equation}
substitution of which into (\ref{Sch}) gives a rise by natural manner to
algebraic relation on parameters $\mu$ and $\lambda$:
\begin{equation}\label{curve}
\mu^2=-\mfrac{1}{2}\,F\,F''+\mfrac{1}{4}\,F'^2+(u+\lambda)\,F^2.
\end{equation}
For direct KdV hierarchy, $F$ is known to be a polynomial on $\lambda$ with
coefficients defined by means of integro-differential operator conjugated to
recursion one.
It occures that $F$ is calculated according to elementary formula:
\begin{equation}\label{F}
\begin{array}{c}
\displaystyle F=\sum_{n=0}^g  \sum_{j=0}^{g-n}\, \lambda^n \,c_j\,F_{g-n-j},
\quad c_0^{}=F_0^{}=1,\quad F_1^{}=-\mfrac{1}{2}\,u,\\
\displaystyle \!\!\!\!\! F_k=\mfrac{1}{8}\,\sum_{j=1}^{k-1} \left(
2\,F''_j\,F_{k-j-1} -F'_j\,F'_{k-j-1} - 4\,F_j\,F_{k-j} - 4\,u\,F_j\,F_{k-j-1}
\right)- \mfrac12\,u\,F_{k-1}\,\,\,(k>1).
\end{array}
\end{equation}
Substituting of (\ref{F}) into (\ref{psi}, \ref{curve}) supplies solution
of spectral problem (\ref{Sch}) with arbitrary finite-gap potential.
It is important that the potential and constants $c_i^{}$ enter Eq.(\ref{psi})
in explicit manner as opposed to well-known Dubrovin's formulae 
\cite{dubrovin}.

{\it Remark.} Analysis of comparatively recently uncovered paper of Drach
\cite{drach2} exhibits that formula (\ref{psi}) was already there.
Although a presence of the potential and constants $c_i^{}$ in this formula 
was realized distinctly by the author, it contains no them explicitly.
It is a wonder that one easily extracts from his formulae the stationary 
equations of Novikov and their nontrivial integrals (in \cite{drach2} they 
were introduced as differential polynomials $d_i$); 
the equations of Dubrovin on roots $\gamma_i(x)$ of polynomial
$F=(\lambda-\gamma_1) \cdots (\lambda-\gamma_g)$ and the first formula of
traces were directly written out and a number of other known just now facts 
were obtained%
\footnote{All these are written on two pages of a text.}.
Moreover, in previous works \cite{drach1}, Drach presented a structure of
solution (\ref{psi}) with function $F$ polynomial on $\lambda$ without
resorting to (\ref{tmp}), relying on own {\em la m\'ethode d'int\'egration 
logique}.
Setting of a problem in \cite{drach2,drach1} is close to one in this report:
{\it with what potentials does one succeed in writing out a solution of the
spectral problem in the quadratures?}

The question originating is how does one seek the formulae like (\ref{psi})
for arbitrary spectral problems or the operator bundles?
Using the results of Krichever \cite{krichever} on an equivalence of the
schemes based on $\Psi$-function as Baker-Akhiezer function and on an 
existence of paired commuting operator \cite{burchnall}, we conclude that the 
last property is not only equivalent to an existence of common eigenfunction 
of two operators and algebraic relation between them, but leads also to 
explicit formula for it.
This fact may be regarded as fundamental and constructive: {\it finite-gap
potentials form simplest class, for which one succeeds in solving the direct
problem, namely, in writing out the formula for $\Psi$
$($non-\mbox{$\,\Theta$}-functional\/$)$}.
At that the property of periodicity (monodromy) is not involved, the
transition to stationary variable is natural, parameter $\mu$ of the variables
separation is an eigenvalue of commuting operator, algebraic curve is {\it a
sequence} of $\Psi$-function formula.

Spectral problem (\ref{Sch}) and treatments (\ref{tmp}--\ref{F}) following it
are only simplest case of general situation.
Next nontrivial example is significant through demonstrating an universality 
of the algorithm of the $\Psi$-function construction by successive elimination
of the derivatives.
Let us consider spectral problem
\begin{equation}\label{sawada}
\Psi'''-u\,\Psi'=\lambda\,\Psi.
\end{equation}
An example of the operator bundle commuting with (\ref{sawada}) is deduced,
to take an instance, from $[L,A]$-pair for Sawada--Kotera equation
\begin{equation}\label{sk}
u_t=u_{\mbox{\scriptsize\it xxxxx}} - 5\,(u\,u_{\mbox{\scriptsize\it xxx}}
+u_x\, u_{\mbox{\scriptsize\it xx}}
-u^2\,u_x)
\end{equation}
after introducing stationary variable $x\to x-\alpha\,t$ and has form:
\begin{equation}\label{stat}
-3\,(u'+3\,\lambda)\,\Psi''+(u''+u^2+\alpha)\,\Psi'+
6\,\lambda\,u\,\Psi=\mu\,\Psi.
\end{equation}
Eliminating all derivatives of $\Psi$ from (\ref{sawada}, \ref{stat}), we 
obtain the algebraic curve.
As it was remained unnoticed, next to the last step in an elimination gives a 
rise to the $\Psi$-formula:
$$
\begin{array}{l}
\Psi(x;\lambda)
=\exp\! {\displaystyle\int\!\mfrac {
\mu\,(2\,u''-v)-3\,\lambda\,(27\,\lambda^2+
2\,v''-2\,v\,u-7\,u'^2)}
{3\, (3\,\lambda+u' )\,u'''
-3\,(\mu+6\,\lambda\,u )\,u'-9\,\lambda\,(\mu+3\,\lambda\,u)
-(2\,u''+v)\,u''-3\,u\,u'^2+v^2}}\,\mbox{\footnotesize $dx$}
\end{array}
$$
$(v\equiv u^2+\alpha)$, from which, to the point, one easily derives an 
expression for product $\Psi_1 \Psi_2 \Psi_3$ being meromorphic function on 
the curve.

A good illustration is supplied by unexpected solution of stationary equation 
(\ref{sk})
$$
u=6\,\wp_1^{}+6\,\wp_2^{}, \quad
\wp_1^{} \equiv \wp(x-\alpha\,t-\Omega;\,g_2^{},g_3^{}), \quad
\wp_2^{} \equiv \wp(x-\alpha\,t-\tilde\Omega;\,g_2^{},\tilde g_3^{}), \quad
\alpha=-12\,g_2^{},
$$
found by Chazy in 1910 \cite[p.380]{chazy}.
Its $\Psi$-function and spectral characteristics cannot be obtained in the 
frameworks of the theory of elliptic solitons.
The product mentioned above will be a polynomial on $\lambda$
$$
\Psi_1 \Psi_2 \Psi_3=\lambda^4+4\,(\wp'_1+\wp'_2)\,\lambda^3+
16\,\wp'_1\,\wp'_2\,\lambda^2+16\,(g_3^{}-\tilde g_3^{})
(\wp'_1-\wp'_2)\,\lambda
-16\,(g_3^{}-\tilde g_3^{})^2,
$$
whose number of zeros coincides, as it should be, with a genus of 
corresponding curve
$$
\mu^3-324\,g_2^{}\,\lambda^2\,\mu+729\,\lambda\,
\Big((\lambda^2+4\,g_3^{}+4\,\tilde g_3^{})^2-64\,g_3^{}\,\tilde g_3^{}\Big)=0.
$$

Other consequence of $\Psi$-formula is a possibility to produce the algebraic
curves in terms of the coefficients of an expansion of the potentials.
In the case of Eq.(\ref{Sch}), constants $c_i^{}$ entering (\ref{curve}) are 
defined from stationary Novikov's equations.
This procedure is elementary at bottom, but there is no necessity even in it
as far as all information is contained in the equation on curve (\ref{curve}).
As it follows from this equation, arbitrary $g$-gap potential of equation 
(\ref{Sch}) can have polar expansions of form 
$$
u(x)=\mfrac{A}{(x-x_0^{})^2}+a_0^{}+a_1^{}\,(x-x_0^{})+\cdots,\qquad
A=2,\,6,\ldots,\,g(g+1).
$$
Substituting this formula in (\ref{curve}) and equalizing to zero a principal 
part of Laurent expansion at neighborhood of {\it any} pole, we define
constants $c_i^{}$.
Next term of the expansion presents the curve equation through remaining free 
coefficients $a_i^{}$.
Also such trick operates in general case, allowing one to solve important 
practical problem of writing out the formulae for the algebraic curves.
In particular, there exists a large number of results of classification
nature in the elliptic solitons theory, but a main difficulty is to find the 
spectral characteristics of the potentials, i.e. the formulae for the curves 
and the coverings of torus.
Here they are obtained as direct consequence of formulae (\ref{psi}) and 
(\ref{curve}).
Thus, in the genus $g=2$ case, we come to natural completion of one of the 
results of paper \cite{enolskii}.

{\bf Proposition:} \it Let $u(x)$ be arbitrary 2-gap potential of equation 
$(\ref{Sch})$.
Then $\Psi$-function is given by expression
$$
\Psi(x;\lambda)=\sqrt{F\,}\,\exp\!\int\!\mfrac{\mu}{F}\,
\mbox{\footnotesize $dx$},
\qquad
F=\lambda^2+\left(c_1^{}-\mfrac{u}{2}\right) \lambda+\mfrac{3}{8}\,u^2-
\mfrac{1}{8}\,u_{\mathit{xx}}-\mfrac12\,\,c_1^{}\,u+c_2^{}.
$$
Under normalization $x_0^{}=a_0^{}=0$, for two possible expansions of the 
potential
$$
u=\mfrac{6}{x^2} + a\, x^2+b\,x^4+c\,x^6+d\,x^8+
\cdots \qquad (c_1^{}=0,\,\, 4\,c_2^{}=-35\,a),
$$
$$
u=\mfrac{2}{x^2} + a\,x^2+b\,x^3+c\,x^4-
\mfrac{3}{10}\,c_1^{}\,b\,x^5+d\,x^6+\cdots \qquad (4\,c_2^{}=-5\,a),
$$
corresponding algebraic curves have form:
$$
\mu^2=\lambda^5-\mfrac {35}{2}\,a\,\lambda^3+\mfrac {63}{2}
\,b\,\lambda^2+\mfrac {27}{8} \left (21\,{a}^{2}+22\,c\right)
\lambda-\mfrac {1377}{4}\,a\,b+\mfrac {1287}{2}\,d,
$$
$$
\begin{array}{ccl}
\mu^2&=& \displaystyle	\lambda^5+2\,c_1^{}\,\lambda^4+
\left(c_1^2-\mfrac52 \,a\right)
\lambda^3-\left (5\,a\,c_1^{}+\mfrac{7}{2}\,c\right)_{\mathstrut} \lambda^2-
\\&& \displaystyle
-\left( \mfrac{5^{\mathstrut}}2\,a\,c_1^2+7\,c\,c_1^{}-\mfrac{27}{8}\,a^2+
\mfrac{81}{4}\,d\right )
\lambda  - \mfrac72\,c\,c_1^2+{\mfrac {27}{8}} \left(a^2-6\,d\right ) c_1^{}+
\mfrac {81}{64}\,b^2.
\end{array}
$$
\rm
It is not difficult to obtain analogous formulae for the higher genus 
potentials.

The potentials for (\ref{sawada}, \ref{stat}) split in 2 series:
$u=(6,12)\,x^{-2}+\cdots$.
Trigonal curve has generically 10 finite branch points $\lambda_i^{}$ with 
indices $(2, 1)$ and a branching at infinity with index 3.
Genus is $g=4$.
We bring expressions only for a case, which can be written in compact manner:
$$
u=\mfrac{12}{x^2}+a\,x^2+b\,x^3+c\,x^4+\left(
\mfrac{a^2}{36}\,x^6+\mfrac{b\,a}{22}\,
x^7+\mfrac{b^2+a\,c}{44}\,x^8+\mfrac {5\,b\,c}{156}\,x^9\right)
+d\,x^{10}+\cdots
\,\,\, (\alpha=-20\,a),
$$
$$
\mu^3-36\,(15\,a\,\lambda^2+49\,b^2)\,\mu+9\,
\lambda\,(3^4\,\lambda^4+3024\,c\,\lambda^2-1568\,a^3 - 2^6\, 21^2\,c^2 +
2^8 \,3^3 \, 637\,d) = 0.
$$

A problem of generalization of the Drach--Dubrovin equations and the formulae 
of traces will be considered in a separate paper. 
Authors are thankful to Prof.~Tsarev~S.\,P. for the discussions of Drach's 
ideas and to Korablinova Nina for the literature supply (especially for the 
copies of articles of Jules Drach). 
This work is partially supported (B.\,Yu.\,V.) by RFBR grant $\#$ 00-01--00782.


\begin{thebibliography}{99}

\bibitem{novikov} Novikov S.P. {\em Funktsional'nyi Analiz i Pril.\/} (1974),
{\bf 8}(3), 54-66 (in Russian).

\bibitem{krichever} Krichever I.M. {\em Usp. Mat. Nauk\/} (1977), {\bf 32}(6),
183-208 (in Russian).

\bibitem{dubrovin} Dubrovin B.A. {\em Funct. Anal. Appl.\/} (1975), {\bf 9}, 
215-223.

\bibitem{drach2} Drach M.J. {\em Compt. Rend. Acad. Sci.\/}
(1919), {\bf 168}, 337-340.

\bibitem{drach1} Drach M.J. {\em Compt. Rend. Acad. Sci.\/}
(1918), {\bf 167}, 743-746. (1919), {\bf 168}, 47-50.

\bibitem{burchnall} Burchnall J.L. and Chaundy T.W. {\em  Proc. L. Math.
Soc.\/}(2) (1922), {\bf 21}(1435), 420-440.

\bibitem{chazy} Chazy J. {\em Acta Math.\/} (1911), {\bf 34}, 317-385.

\bibitem{enolskii} Belokos E.D. and Enolskii V.Z. {\em Funktsional'nyi Analiz
i Pril.\/} (1989), {\bf 23}(1), 57-58 (in Russian).

\end{thebibliography}
\end{document}